\documentstyle[aps,multicol,epsfig]{revtex}

\begin{document}

\title{Observation of strong surface state effects in the nonlinear magneto-optical response of Ni(110)}

\author{K. J. Veenstra$^1$, A. V. Petukhov$^{1,2,*}$ , E. Jurdik$^1$ and Th. Rasing$^1$}
\address{$^1$Research Institute for Materials, University of Nijmegen, 
Toernooiveld 1, 6525 ED Nijmegen, The~Netherlands}
\address{$^2$SON Research Institute, University of Nijmegen, 
Toernooiveld 1, 6525 ED Nijmegen, The~Netherlands}
\maketitle

\begin{abstract}
Spectroscopic magnetization induced optical Second Harmonic Generation (MSHG) 
measurements from a clean Ni(110) surface reveal strong resonance effects near 
2.7~eV that can be attributed to the presence of an empty surface state. 
The good agreement with model calculations shows the potential of MSHG to probe 
spin polarized interface band structures.

\end{abstract}


\begin{multicols}{2}
 
The spin dependent electronic structure of 
ferromagnetic surfaces and interfaces forms the fundamental basis for 
understanding magnetic phenomena like giant magneto 
resistance (GMR) and interlayer exchange coupling, that have
attracted significant attention also because of their technological 
importance. 
Magnetic Second Harmonic Generation (MSHG) is a new magneto-optical tool 
that is intrinsically sensitive to the structure and magnetism of surfaces and 
interfaces \cite{Rasing99}. After its introduction in 
1991 \cite{Rei91}, 
MSHG has been used to study magnetic properties of e.g. 
iron \cite{Vollmer96-2}, polycrystalline nickel\cite{Boeh95}, 
magnetic multi-layer structures (where buried layers were probed)\cite{Wier95-2,Kirilyuk96,Kirilyuk98}, 
and to study the ultrafast magnetization dynamics in 
ferromagnets \cite{Hohlfeld97,Knorren99}.
As the MSHG response, on a microscopic level, involves coherent electronic  
transitions between occupied and unoccupied electronic states in the 
spin-polarized bandstructure \cite{Hueb90,Hueb94,Pus93}, MSHG is  
an ideal spectroscopic tool to probe the spin-polarized 
density of states at ferromagnetic surfaces and interfaces.

In this paper we present a phase sensitive spectroscopic MSHG 
(SH-energy 2.4~eV - 3.3~eV) investigation of a clean Ni(110) surface
that shows that the MSHG response is sensitive to transitions from exchange 
split d-bands into empty surface states. The spectroscopic measurements were
phase resolved  and we show that this unique feature is essential to obtain 
a correct interpretation of results in the vicinity of resonances.
These transitions occur around 2.7~eV  
and lead to a maximum in the measured magnetic asymmetry and an exchange
splitting in the magnetic tensor components. A simple model, that 
incorporates the spin dependent density of states, gives an excellent 
description of the observed effects.


The second harmonic polarization $P(2\omega)$ induced by a fundamental 
laser field $E(\omega)$ can be written as
\begin{equation}
\label{SHpol} 
P_i(2\omega) = \chi^{\rm (2)}_{ijk}(-2\omega;\omega,\omega)E_{j}(\omega)E_{k}(\omega)
\end{equation}
with $\chi^{(2)}$ being the dipole second order nonlinear susceptibility 
tensor. In centrosymmetric materials this tensor is only nonzero at surfaces 
or interfaces where the inversion symmetry is broken\cite{Shenbook}.
The presence of a magnetization ${\bf M}$ breaks the time reversal symmetry 
and introduces odd tensor elements $\chi^{\rm odd}_{ijk}$ \cite{Rasing99}. 
These tensor elements change sign upon magnetization reversal and therefore
give rise to the magnetic asymmetry in the MSHG response. 
On a microscopic level this asymmetry arises like in linear magneto optics due
to the splitting of the energy levels by both the exchange and 
the spin orbit interaction \cite{Arg55,Bruno96}.
The relation between the microscopic spin dependent bandstructure and the
nonlinear magneto optical susceptibility,
can be written as
\begin{eqnarray}
\label{response}
\chi^{(2)}_{ijk}(-2\omega;\omega,\omega)\propto \\  \sum_{a,b,c} 
\frac{<a|i|c><c|j|b><b|k|a>}{(2\hbar \omega- E_{ca}+i\hbar \Gamma_{ca})
(\hbar \omega- E_{ba}+i\hbar \Gamma_{ba})}\nonumber
\end{eqnarray}
where $|a>$, $|b>$ and $|c>$ are spin dependent  initial,  intermediate and 
final states. 
The measured intensity in a fixed experimental geometry with opposite 
magnetization directions $\pm {\bf M}$, 
can in general be written as a sum of effective tensor components:
\begin{equation}
\label{SHint} 
I^{\pm}(2\omega)  \propto  |\chi^{\rm even}_{\rm eff}(2\omega) 
\pm \chi^{\rm odd}_{\rm eff}(2\omega)|^2, 
\end{equation} 
where $\chi^{\rm even}_{\rm eff}$ and $\chi^{\rm odd}_{\rm eff}$ are linear 
combinations of the even and
odd tensor elements and Fresnel factors $\alpha_{ijk}$:
\begin{equation}
\label{chieff}
\chi_{\rm eff} = \sum_{i,j,k} \alpha_{ijk}\chi_{ijk}.
\end{equation}
The number of nonzero 
tensor elements in the summation in Eq.(\ref{chieff}) depends upon the 
symmetry of the surface and the optical polarization geometry. 
Note that only in the 
case of highly symmetric, e.g. isotropic, surfaces it 
is possible to separate the odd and even contributions in the MSHG response 
by choosing a particular polarization combination\cite{Stolle97}.
In the case of a 
Ni(110)-surface with a magnetization parallel to the 
easy ($\overline{1}$11)-axis, the nonlinear susceptibility tensor  contains 
18 (8 even and 10 odd) independent nonzero elements \cite{Veenstra99}. 
The magnetic asymmetry, as measured in an MSHG measurement can be defined as
\begin{equation}
\label{Asym}
\rho=\frac{I^{+} -I^{-}}{I^{+}+I^{-}}= 
\frac{2| \chi^{\rm odd}_{\rm eff}| /|\chi^{\rm even}_{\rm eff}|}{1+| \chi^{\rm odd}_{\rm eff}/
\chi^{\rm even}_{\rm eff}|^2}\cos(\Delta\Phi).
\end{equation}
with $\Delta\Phi$ the phase difference between the odd and even effective 
tensor components. 
Because the asymmetry $\rho$ is normalized with respect to the
total SH-intensity, $I^{+}+I^{-}$, it does not depend upon
the intensity or shape of the fundamental light pulses, nor on the spectral 
properties of e.g. filters in the optical setup. However, from 
Eq.(\ref{Asym}) it follows that a measurement of only $\rho$ does not suffice 
to determine the spectral dependence of the effective odd and even 
susceptibilities separately.
Therefore the relative phase between  $|\chi^{\rm odd}_{\rm eff}| $ and 
$|\chi^{\rm even}_{\rm eff}|$ should be measured and in addition the 
intensity should be normalized to a reference.


The MSHG experiments were performed at room temperature on a disk shaped
Ni(110) single crystal placed between the poles of an {\it in situ} electromagnet in a 
UHV system with a base pressure of $5 \times 10^{-11}$ mbar.
The sample surface was cleaned by repeated cycles of 550 eV Ar$^+$ sputtering 
and e-beam heating to 1000 K, until no contamination could be traced by 
Auger-electron
spectroscopy and a sharp (1 $\times$ 1) LEED (low energy electron diffraction)
pattern could be observed. For each MSHG measurement one cleaning 
cycle was repeated.
A tunable optical parametric amplifier (OPA) pumped by a Ti-sapphire 
regenerative amplifier was used to produce the fundamental light pulses of 
100 fs duration in the wavelength range 840~nm - 1000~nm with a repetition rate of 1kHz. Between 750~nm and 850~nm the direct output of a 
Ti-sapphire laser (rep. rate 82~MHz) was used.  
The Second Harmonic light from the Ni sample was detected with a photomultiplier 
tube. To normalize the measured SH-intensity from the nickel,
the SH-intensity from a c-cut quartz crystal in the transmission geometry
was measured with a second photomultiplier tube.
Color filters (BG39) were used to filter out the fundamental light.
The phase of the SH-light was measured using the UHV compatible phase 
sensitive detection technique recently developed by us \cite{Veenstra98}.
The latter is based upon interference spectroscopy where the relative 
phase of a sample and reference pulse, delayed with respect to each other by a 
time $\tau$, can be extracted from the spectrum. 

\begin{figure}
\noindent\begin{minipage}{\linewidth}\vspace{0.1\linewidth}
\noindent\hspace*{\fill}\includegraphics[width=0.8\linewidth]{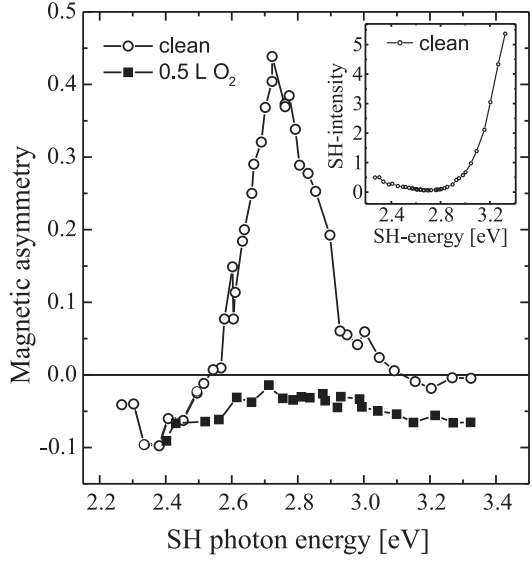}
\hspace*{\fill}\vspace{0.2\linewidth}
\caption{Magnetic asymmetry $\rho$ as a function of Second Harmonic photon energy in  p$_{\rm in}$-p$_{\rm out}$ polarization combination as measured on a 
clean and oxidized Ni(110) surface.
Inset: Average Second Harmonic intensity as a function of Second Harmonic photon energy.}
\label{rho}
\end{minipage}
\end{figure}

\begin{figure}
\noindent\begin{minipage}{\linewidth}\vspace{0.1\linewidth}
\noindent\hspace*{\fill}\includegraphics[width=0.8\linewidth]{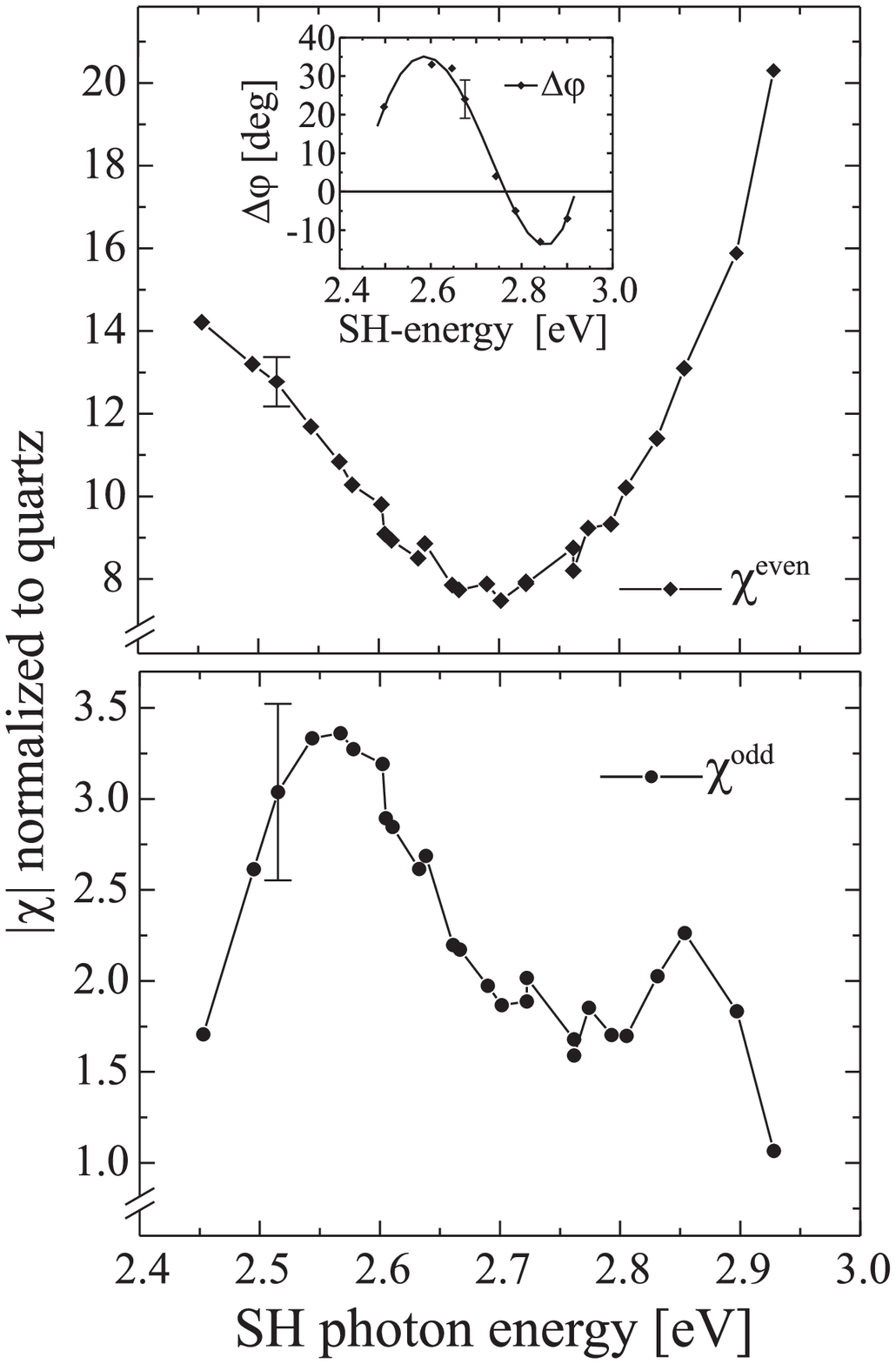}
\hspace*{\fill}\vspace{0.1\linewidth}
\caption{Amplitude of the effective tensor elements 
$\chi^{\rm even}$ and $\chi^{\rm odd}$} as derived from the measured intensity, asymmetry and relative phase $\Delta \varphi$. In the inset the measured 
frequency dependence of $\Delta \varphi$ is plotted.
\label{chis}
\end{minipage}
\end{figure}

The magnetic asymmetry $\rho$ 
measured in the polarization combination p$_{\rm in}$-p$_{\rm out}$ 
is plotted in Fig.(\ref{rho}) as a function of SH-photon energy.
The open circles represent the data measured on the clean surface whereas the
solid squares represent the data measured on the oxidized (0.5~L O$_{2}$) 
surface. In the inset of Fig.(\ref{rho}), the average 
SH-intensity $(I^{+}+I^{-})/2$ as measured on the clean surface 
is shown. $I^{+}$ and $I^{-}$ were measured at zero (applied) field after 
saturating (0.07 T) the Ni-sample along the easy ($\overline{1}$11)-axis. 
The magnetic asymmetry $\rho$ of the clean surface changes sign 
at 3.1~eV and 2.6~eV and has a maximum at 2.7~eV. This 
resonant feature disappears upon oxidation, clearly proving its surface 
specific nature. The relation between the effective susceptibilities and the 
data in Fig.(\ref{rho}) is given by
\begin{equation}
\label{ampchis}
4|\chi_{\rm eff}|^2=I^{+}+I^{-} \pm 
2 \sqrt{I^{+}I^{-}}\cdot \cos(\Delta \varphi),
\end{equation}
where $\Delta \varphi$ is phase difference between 
$E(2\omega,+{\bf M})$ and $E(2\omega,-{\bf M})$. This phase difference 
has been measured as a function of frequency using the technique described in 
\cite{Veenstra98} and is shown in the inset of Fig. (\ref{chis})
The resulting effective susceptibilities 
$|\chi^{\rm odd}_{\rm eff}|$ and $|\chi^{\rm even}_{\rm eff}|$ are also shown 
in Fig.(\ref{chis}).
Clearly $|\chi^{\rm odd}_{\rm eff}|$ has two maxima at 2.82~eV and 2.55~eV 
respectively whereas $|\chi^{\rm even}_{\rm eff}|$ has a minimum at 2.7~eV.
The error in $\chi^{\rm odd}$ as indicated in Fig.(\ref{chis}) essentially
results from the error in $\Delta \varphi$ which is typically 5$^\circ$.

\begin{figure}
\noindent\begin{minipage}{\linewidth}\vspace{0.1\linewidth}
\noindent\hspace*{\fill}\includegraphics[width=0.8\linewidth]{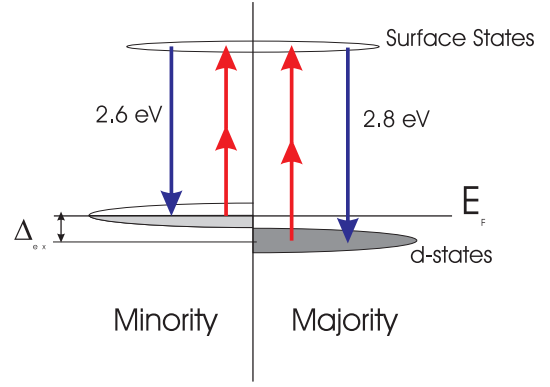}
\hspace*{\fill}\vspace{0.1\linewidth}
\caption{Schematic picture of the exchange split density of states of nickel 
and empty surface states.}
\label{model}
\end{minipage}
\end{figure}

\begin{figure}
\noindent\begin{minipage}{\linewidth}\vspace{0.1\linewidth}
\noindent\hspace*{\fill}\includegraphics[width=0.8\linewidth]{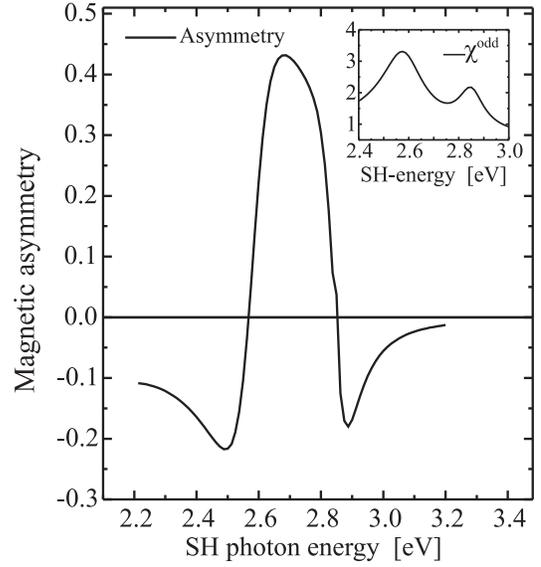}
\hspace*{\fill}\vspace{0.1\linewidth}
\caption{Asymmetry as calculated according to the simple two Lorentzian model 
in Eq.(\ref{lorentz}). Inset: $\chi^{\rm odd}$ according to the same model.}
\label{lor}
\end{minipage}
\end{figure}

The resonances as observed in the nonlinear magneto-optical spectra in 
Figs.(\ref{rho}) and (\ref{chis}) can be explained within a simple model shown 
in Fig.(\ref{model}).
The model involves the spin splitting of the d-bands around the Fermi energy 
and an empty surface state band around 2.5~eV above $E_{\rm F}$.
The exchange splitting of the d-band leads to a 
maximum density of states for minority spin electrons  
at the Fermi energy and a maximum for majority spins approximately
250~meV below $E_{\rm F}$ \cite{Donath94}. Several inverse photoemission 
studies 
\cite{Goldmann85,Donath94,Bertel95} have reported an empty surface state band 
at the 
$\overline {\rm Y}$ point of the fcc (110) surface Brillouin zone. As this surface 
sate is of nearly pure p$_{\rm z}$ character \cite{Bertel95} the exchange 
splitting of this
state is much smaller than the splitting of the d-states and can be neglected.
Including only these d-states and surface states into the summation of 
Eq.(\ref{response}), 
$\chi^{\rm odd}$ can be written as
\begin{eqnarray}
\label{lorentz}
\chi^{\rm odd}={A_0} + \frac{A_1}{2\hbar \omega - E_{1} +i\hbar \Gamma_{1}} + \frac{A_2}{2\hbar \omega - E_{2} +i\hbar \Gamma_{2}} 
\end{eqnarray}
where  the second term describes the transitions of the minority spin 
electrons 
from filled to empty states having energy difference $E_1$ and the third 
term includes the transitions of the majority spin electrons. $A_{1,2}$ 
include the  matrix elements and the nonresonant energy factor  
from Eq.(\ref{response}). $A_0$ is a constant 
background term including all non resonant contributions. Like
in linear optics this is related to an integration of all possible vertical 
transitions over the complete bandstructure.
The widths of the transitions are given by $\Gamma_{1,2}$. 
Because of the spin dependence of the resonant contributions to 
the \emph{odd} 
tensor component, the matrix elements $A_1$ and $A_2$ should have an opposite 
sign and are related according to $A_{1}=-|x| A_{2}$, 
where the factor $|x| \sim 1$ accounts for the 
possible difference in their absolute values. 
Note that in our model both $A_{1,2}$ and $A_0$ may be complex; however 
for the fitting, only  
the relative phase of  $\chi^{\rm odd}$ is relevant and is determined by 
taking $A_0$ complex and $A_{1,2}$ real. 
Using the relation between $A_1$ and $A_2$ we can now fit the 
model to the data in Fig.(\ref{chis}) and obtain $\chi^{\rm odd}$ as shown in 
the inset of Fig.(\ref{lor}) with $E_{1}=2.58$~eV and $E_{2}=2.85$~eV.
Once also $\chi^{\rm even}$ is known, it is possible to check the 
model by calculating $\rho$ according to Eq.(\ref{Asym}). 
However, $\chi^{\rm even}$ cannot simply be described by only the transitions 
shown in Fig.(\ref{model}). We assume that that the behavior of 
$\chi^{\rm even}$ is mainly determined by non-resonant contributions and 
partially by the resonances as indicated in Fig.(\ref{model}).
Because of this large 
non-resonant background its relative
phase does not change as much as the relative phase 
of  $\chi^{\rm odd}$. Therefore we can fit $\chi^{\rm even}$ simply to 
a real fourth rank polynomial. 
With this the magnetic asymmetry can now be calculated according to 
Eq.(\ref{Asym}) and is
shown in Fig.(\ref{lor}). The typical features of the asymmetry like the two 
sign changes and the maximum are described very well by the model. 
Although the fit to $\chi^{\rm odd}$ in Fig. (\ref{lor}) is not unique, 
to get the 
agreement the important physical constraint in the model is 
$A_{1}=-|x| A_{2}$, i.e., the two resonances must have opposite phases.
If we take $A_{1}=+|x| A_{2}$ it is also 
possible to fit $\chi^{\rm odd}$, however one does not obtain a reasonable 
agreement for the magnetic asymmetry. This indicates that the features in 
$\rho$ do arise due to the difference in exchange splitting between the 
initial (d-states) and the final (surface) states
 which proves that MSHG-spectroscopy in general can be a
powerful tool to probe the spin-dependent electronic structure of surfaces 
and also buried interfaces.


In summary, we have measured a resonance in the nonlinear magneto-optical 
response from a clean Ni(110) surface in a UHV environment. This resonance 
disappears very rapidly upon oxidation of the surface (at 0.5~L)
indicating its surface state origin. By performing phase 
sensitive MSHG-spectroscopy we have determined the effective susceptibility 
spectra that can be fitted with a simple model involving the 
exchange splitting of d-states at and below the Fermi energy and an empty 
surface state above the Fermi energy. According to our model the exchange splitting
of the surface state should be much smaller than the splitting of the d-states, which confirms the p-like character of this surface state. Our results show 
that MSHG spectroscopy can indeed probe spin-dependent interface 
bandstructures. It will be very interesting to apply this technique to buried 
interfaces such as that between a ferromagnet and a tunnel barrier. With 
typical film thicknesses of a few nm, this should in principle be no problem.

Part of this work was supported by the Stichting Fundamenteel Onderzoek der 
Materie (FOM) and by the TMR network NOMOKE. Skillful assistance of 
A.F. van Etteger is gratefully acknowledged.

\newpage

\end{multicols}
\end {document}